\documentclass[12pt]{iopart}
\usepackage{iopams}
 \usepackage{amsfonts}
\usepackage{graphicx}
\begin{document}
\title{Imbalanced superfluid state in an annular disk}

\author{Fei Ye$^{1,3}$, Yan Chen$^{2}$, Z D Wang$ ^{3}$ and F C Zhang$^{3}$}

\address{$ ^{1}$Center for Advanced Study, Tsinghua University, Beijing,
  100084, China}

\address{$ ^{2}$Department of Physics and Lab of Advanced Materials,
  Fudan University, Shanghai, 200433, China}

\address{$ ^3$ Department of Physics and Center for Theoretical and
  Computational Physics, The University of Hong Kong, Hong Kong,
  China}

\eads{\mailto{feiye@mail.tsinghua.edu.cn},
      \mailto{yanchen99@fudan.edu.cn},
      \mailto{zwang@hkucc.hku.hk},
      \mailto{fuchun@hkucc.hku.hk}}

\begin{abstract}
  The imbalanced superfluid state of spin-1/2 fermions with $s$-wave
  pairing is numerically studied by solving the Bogoliubov-de-Gennes
  equation at zero temperature in an annular disk geometry with narrow
  radial width. Two distinct types of systems are considered. The
  first case may be relevant to heavy fermion superconductors, where
  magnetic field causes spin imbalance via Zeeman interaction and the
  system is studied in a grand canonical ensemble.  As the magnetic
  field increases, the system is transformed from the uniform
  superfluid state to the Fulde-Ferrell-Larkin-Ovchinnikov state, and
  finally to the spin polarized normal state. The second case may be
  relevant to cold fermionic systems, where the numbers of fermions of
  each species are fixed as in a canonical ensemble. In this case, the
  groundstate depends on the pairing strength. For weak pairing, the
  order parameter exhibits a periodic domain wall lattice pattern with
  a localized spin distribution at low spin imbalance, and a
  sinusoidally modulated pattern with extended spin distribution at
  high spin imbalance.  For strong pairing, the phase separation
  between superfluid state and polarized normal state is found to be
  more preferable, while the increase of spin imbalance simply changes
  the ratio between them.
\end{abstract}

\pacs{67.85.-d, 03.75.Ss, 74.81.-g, 74.25.Ha}
\submitto{\JPC}

\maketitle

\section{Introduction}
In a conventional BCS theory, the normal state has a Fermi surface
common to both spin-up and spin-down electrons and the Cooper pair has
a zero total momentum.  More than forty years ago, Fulde and
Ferrell\cite{fulde1964}(FF), Larkin and Ovchinnikov\cite{larkin1965}
(LO) proposed independently the pairing mechanism for the mismatched
Fermi surfaces due to the spin imbalance. In the FF state, a spin up
electron with momentum $\vec{k}$ is bounded with a spin down electron
with momentum $-\vec{k}+\vec{q}$, thereby the Cooper pair has a net
momentum $\vec{q}$ which is determined by the imbalance between two
Fermi surfaces. Therefore the order parameter is characterized by a
single momentum $\vec{q}$, which can be written as
$\Delta(\vec{r})=\Delta_0 e^{i\vec{q}\cdot \vec{r}}$ with a uniform
magnitude $\Delta_0$. If considering the composition of two momenta,
$\vec{q}$ and $-\vec{q}$, one gets the LO state where the order
parameter is real with its magnitude oscillating periodically in
space. 

In condensed matter physics, the spin imbalance can be generated by
applied magnetic fields. However the condition for the FFLO state to
be observed is quite stringent on the superconducting
materials. Roughly speaking, there are three requirements (i) low
$T_c$, so that the magnetic field needed to imbalance the spin
population is accessible; (ii) the orbital effect of magnetic field is
weak enough to avoid pair breaking before the Zeeman splitting takes
effect; (iii) clean limit, i.e., the mean free path of electron should
be much longer than the correlation length, since the FFLO state is
easily destroyed by impurities. Some of heavy fermion superconductors
are good candidates to fulfill these requirements (for a review see
Ref.~\cite{matsuda2007}). There was recent indications that
CeCoIn$_{5}$ indeed exhibits the FFLO
state\cite{radovan2003,bianchi2003,capan2004,watanabe2004,miclea2006,kumagai2006}.
That compound is a quasi-two-dimensional heavy fermion superconductor
with a $d$-wave pairing. In the cold fermionic atom system with
different hyperfine spins, the spin population imbalance between
different hyperfine spins can be easily controlled by applying radio
frequency field. Recently the imbalanced superfluid state has been
realized in these cold neutral atom systems
\cite{zwierlei2006a,zwierlein2006b,partridge2006a,partridge2006b,shin2006},
and the possible spatially modulated superfluid phases in these
systems are studied in Ref. \cite{mizushima2005,machida2006}. It is
noted that the particle number of different species may be controlled 
directly in systems like cold atoms, and in superconductors the spin
imbalance is generated by the external magnetic fields, which may
correspond to two different thermodynamic conditions, respectively.

In a recent theoretical study~\cite{chen2007}, it was found that in
the harmonically trapped polarized fermionic atoms in a
two-dimensional (2D) optical lattice, the insulating core is
surrounded by a superfluid shell at high atom densities with pairing
parameter modulated in the circumferential direction.  Since some of
important physics may be explained by the quasi-one-dimensional
(quasi-1D) shell, it is thus interesting to study further the FFLO
with more details in a quasi-1D system. The possible angular FFLO
state in a toroidal trap has also been investigated in a very recent
study~\cite{yanase2009}. In the present paper, we consider a quasi-1D
annular disk with narrow enough radial width, so that the radial
modulation of the order parameter might result in a quite large radial
gradient of order parameter which increases the system energy
considerably according to the Ginzburg Landau(GL) theory. Therefore
the oscillation of pairing amplitude is suppressed in radial
direction, and restricted only in circumferential direction. In a
large 2D system, the order parameter oscillation has more freedom and
can happen in arbitrary directions. In the presence of inhomogeneity
the modulation direction may vary in space which leads to irregular
pattern of order parameter. Therefore it may be easier to observe
regular oscillations of the pairing amplitude in a quasi-1D system
than in the 2D film.

In this paper, we consider two distinct systems.  The first one may be
relevant to heavy fermion superconductors, where the electrons spins
interact with an external magnetic field via the Zeeman coupling. The
second system may be related to the cold fermionic atoms, where the
number of fermions of each spin is fixed. We employ a grand canonical
ensemble to study the first system and a canonical ensemble to study
the second system.  We solve the Bogoliubov-de-Gennes (BdG) equation
at zero temperature numerically for the above quasi-1D systems. Our
main results can be summarized below.  In the first case, as the
magnetic field increases, the ground state is transformed from a
uniform superfluid state to the sinusoidally modulated LO state, and
then to a spin polarized normal state. In the second case, the ground
state depends on the pairing strength. For weak interactions, the
order parameter exhibits a periodic domain wall lattice pattern with a
localized spin distribution for low spin imbalance, and a sinusoidally
modulated pattern with extended spin distribution for high spin
imbalance.  For strong interactions, the phase separation between
superfluid state and polarized normal state is found to be more
preferable, while increase of spin imbalance simply extends the
spatial region of the normal state. The paper is organized as
follows. In Sec. II, we study the exact 1D case.  In Sec. III, we
present our results for annular disk geometry.  The conclusion is
given in Sec. IV.

\section{Imbalanced Superfluid State in One-dimensional Ring}
\subsection{One-dimensional BdG Equation}

Before exploring the properties of imbalanced superfluid in the
annular disk geometry, we first consider the 1D ring, which may be
viewed as the limiting case where the disk width is so narrow that
only one radial mode is relevant. This case has been studied by a
number of authors.  In the mean field(MF) level, a rigorous analysis
for the 1D BdG equation is given in Ref.~\cite{machida1984} in
the presence of a magnetic field. In terms of 1D Luttinger liquid
theory the imbalanced superconducting state is also elucidated by
Yang~\cite{Yang2001}, and very recently, the density matrix
renormalization group algorithms are implemented on the 1D
negative-$U$ Hubbard model to explore the FFLO state in
Refs.~\cite{feiguin2007,rizzi2008,tezuka2008,feiguin2009}. The
cold fermionic gases with attractive interaction and population
imbalance are studied theoretically in Ref.~\cite{orso2007} and
and Ref.~\cite{hu2007}. 

In this subsection, we follow the MF treatment to give a brief
description to the 1D imbalanced superfluid state.  We consider a
canonical ensemble and fix the number of fermions of different
species.  Although only the quasi-long range order may exist in 1D
system, the MF approach presented in this section is helpful to
understand the imbalanced superfluid in 2D annular disk geometry shown
in later sections.

The mean field Hamiltonian for a 1D interacting system reads
\begin{eqnarray}
\label{eq:1}
\hat{H}&=& \int dx [\sum_{\alpha}\hat{\psi}^{\dagger}_{\alpha}(x)\left(-\frac{\hbar^2\partial^2_x}{2m}
  \right) \hat{\psi}_{\alpha}(x) \nonumber\\
&&+ \left(\Delta(x)\hat{\psi}^{\dagger}_{\uparrow}(x)\hat{\psi}^{\dagger}_{\downarrow} (x)+
h.c.\right)-\frac{|\Delta(x)|^2}{g}]
\nonumber\\
&& - \sum_{\alpha}\mu_{\alpha}[\int dx
\hat{\psi}^{\dagger}_{\alpha}(x) \hat{\psi}_{\alpha}(x)-N_{\alpha}] \nonumber\\
\Delta(x) &=& g \left\langle \hat{\psi}_{\downarrow}(x)
  \hat{\psi}_{\uparrow}(x)  \right\rangle\;.
\end{eqnarray}
$\hat{\psi}_{\alpha}(x)$ is the fermion annihilation field at position
$x$ with spin index $\alpha$, $\Delta(x)$ is the fermion pairing
field, $m$ is the mass of the particle, and $g<0$ is the attractive
interaction strength. $\mu_{\alpha}$ are the Lagrangian multipliers or
the chemical potentials, which are used to fix the numbers of fermions
of different spins at $N_{\uparrow}$ and $N_{\downarrow}$,
respectively.

Eq.~(\ref{eq:1}) has the similar form to the well known
Su-Schrieffer-Heeger(SSH) model \cite{su1979} for polyacetylene, which
describes a 1D electron system coupled to phonons. In this system,
when the phonon fields are condensed in opposite phases at the two
ends of the 1D string, there are possible soliton excitations with
zero energy in the fermion spectrum.  The soliton excitations are also
possible in the 1D superfluid Hamiltonian Eq.~(\ref{eq:1}), where the
MF pairing parameter $\Delta(x)$ can mimic the phonon field in the SSH
model, which is shown briefly below.  More details can be found, e.g.,
in Ref.~\cite{machida1984}.  For simplicity we take
$\mu_{\uparrow}=\mu_{\downarrow}=\mu$, which determines the Fermi
momentum $k_F= \sqrt{2m\mu}/\hbar$. The low energy physics is
described by quasiparticles around the two Fermi points $\pm k_F$,
i.e., the following decomposition is allowed
\begin{eqnarray}
\label{eq:3}
\hat{\psi}_{\sigma}(x) \sim
e^{ik_Fx}\hat{R}_{\sigma}(x)  + e^{-ik_Fx} \hat{L}_{\sigma}(x)
\end{eqnarray}
with left and right movers defined as
\begin{eqnarray}
\label{eq:4}
&&\hat{R}_{\sigma}(x) = \sum_{-\Lambda<k<\Lambda}
\hat{\psi}_{\sigma}(k+k_{F})\frac{e^{ikx}}{\sqrt{L}}  \nonumber\\
&&\hat{L}_{\sigma}(x) = \sum_{-\Lambda<k<\Lambda} \hat{\psi}_{\sigma}(k-k_{F})
\frac{e^{ikx}}{\sqrt{L}} \;.
\end{eqnarray}
$\Lambda$ is a suitable momentum cutoff.  These quasiparticle
operators satisfy the standard anti-commutation relations, i.e.,
\begin{eqnarray*}
\{\hat{R}_{\sigma}, \hat{R}_{\sigma'}^{\dagger}\} = 
\{\hat{L}_{\sigma}, \hat{L}_{\sigma'}^{\dagger}\} = \delta_{\sigma,\sigma'}
\end{eqnarray*} 
and all the other anti-commutators are zero. Substituting
Eq.~\ref{eq:3} into Eq.~\ref{eq:1}, and neglecting the fast
oscillation terms ($\propto\exp(\pm 2ik_Fx)$), one obtains the
following two Hamiltonians to the linear order of $k$,
\begin{eqnarray}
\label{eq:5}
\hat{H}_1&=& \hbar v_F\int dx :\hat{R}^{\dagger}_{\uparrow}(-i\partial_x)
\hat{R}_{\uparrow}:-:\hat{L}^{\dagger}_{\downarrow} (-i\partial_x) \hat{L}_{\downarrow}:
\nonumber\\
&& +\int dx\Delta(x)(\hat{R}^{\dagger}_{\uparrow}\hat{L}^{\dagger}_{\downarrow}+
\hat{L}_{\downarrow}\hat{R}_{\uparrow}) \nonumber\\
\hat{H}_2&=& \hbar v_F\int dx :\hat{R}^{\dagger}_{\downarrow}(-i\partial_x)
\hat{R}_{\downarrow}:-:\hat{L}^{\dagger}_{\uparrow} (-i\partial_x) \hat{L}_{\uparrow}:
\nonumber\\
&& +\int dx\Delta(x)(\hat{L}^{\dagger}_{\uparrow}\hat{R}^{\dagger}_{\downarrow}+
\hat{R}_{\downarrow}\hat{L}_{\uparrow}) \;.
\end{eqnarray}
Here $:A:$ denotes normal ordering of $A$ and $v_F$ means the positive
Fermi velocity. In the following $\hbar v_F$ is taken as
unit. $\hat{H}_1$ and $\hat{H}_2$ are commutative with each other, and
connected through the gap equation
\begin{eqnarray}
\label{eq:2}
\Delta(x)= g \left\langle R_{\downarrow}L_{\uparrow} + L_{\downarrow}
  R_{\uparrow} \right\rangle.
\end{eqnarray}
The order parameter $\Delta(x)$ is assumed to be real. Eq.~\ref{eq:2}
shows that the pairing takes place either between $\hat{L}_{\uparrow}$
and $\hat{R}_{\downarrow}$, or between $\hat{L}_{\downarrow}$ and
$\hat{R}_{\uparrow}$. Actually, $ \left\langle
  R_{\downarrow}L_{\uparrow} \right\rangle = \left\langle
  L_{\downarrow}R_{\uparrow} \right\rangle$ by symmetry. Formally, one
may have $\hat{H}\sim \hat{H}_1 + \hat{H}_2 -\int dx
|\Delta(x)|^2/g$, but it is emphasized that $\hat{H}_{1,2}$ only
describe the low energy excitations near the Fermi surface.

Let's consider only $H_1$ with a twisted $\Delta(x)$, i.e.,
$\Delta(-\infty)=-\Delta(\infty)=\Delta_0$. As shown by Jackiw and
Rebbi\cite{jackiw1976}, there is at least one zero mode
$\hat{\gamma}_{0\uparrow}$ in the middle of the gap, which is
localized in space and reads
\begin{eqnarray}
\label{eq:6}
&&\hat{\gamma}_{0\uparrow} \propto \int dx F(x)[\hat{R}_{\uparrow}(x) - i
 \hat{L}_{\downarrow}^{\dagger}(x)] \nonumber\\
&&F(x)\propto \exp\left[\int^x_0dx'\Delta(x')\right]
\end{eqnarray}
It is easy to verify the commutation relation
$[\hat{\gamma}_{0\uparrow},\hat{H}_1]=0$.  Besides this localized zero
mode, we also have other quasiparticle excitations
$\hat{\gamma}_{n\alpha}$ in the continuum region, where $n$ and
$\alpha$ are the energy level and spin indices, respectively. Assuming
all of them constitute a complete representation of the Hamiltonian
$H_1$, the lowest energy states are doubly degenerate in the presence
of an order parameter with kink pattern, which is the spinless vacuum
of the quasiparticles $\hat{\gamma}_{n\alpha}$ together with the zero
mode $\hat{\gamma}_{0\uparrow}$ being either filled or empty. Similar
analysis is also valid for the $H_2$ branch, for which one can find
that the zero mode has the form 
\begin{eqnarray*}
\hat{\gamma}_{0\downarrow} \propto
\int dx F(x)[\hat{R}_{\downarrow}(x) + i
\hat{L}_{\uparrow}^{\dagger}(x)] 
\end{eqnarray*}
which satisfies $[\gamma_{0\downarrow},H_2] = 0$.

In terms of $\hat{R}_{\sigma}$ and $\hat{L}_{\sigma}$, the total
particle number $\hat{N}$ and total spin operator $\hat{S}$ can be
written as
\begin{eqnarray*}
\label{eq:15}
&& \hat{N} = \hat{N}_{\uparrow} + \hat{N}_{\downarrow}, \hspace{1cm}
\hat{S} = \hat{N}_{\uparrow} - \hat{N}_{\downarrow} \nonumber \\
&&\hat{N}_{\sigma}  = \int dx [:\hat{R}^{\dagger}_{\sigma}
\hat{R}_{\sigma}:+ :\hat{L}^{\dagger}_{\sigma} \hat{L}_{\sigma}:]\;,
\end{eqnarray*}
where the fast oscillating terms are neglected.  Note that the
quasiparticle operators $\hat{R}_{\sigma}$ and $\hat{L}_{\sigma}$ can
only describe the low energy physics, hence the operator
$\hat{N}_{\sigma}$ with normal ordering only measures the particle
number relative to the Fermi surface. Obviously, unlike the SSH
model\cite{su1979} and the Jackiw-Rebbi model\cite{jackiw1976}, the
charge conservation is broken in the BCS theory, therefore one can not
tell how many charges the soliton can carry. Despite this fact, the
total spin is still a conserved quantity in our MF treatment,
therefore each zero mode may carry half spin as an analog to the half
charge investigated in Ref.\cite{su1979,jackiw1976}. But in practice
only one spin can be observed at the kink of $\Delta(x)$, since there
are two branches of fermions($H_1$ and $H_2$). To observe the half
spin, one must get rid of the fermion doubling problem. Nevertheless,
this provides a mechanism to accommodate excess spins with zero
energy. The total energy of the soliton measured relative to the
uniform BCS state is computed to be
$2\Delta_0/\pi$\cite{Dashen1975,Takayama1979}, which is less than the
superfluid gap.

\subsection{From Soliton Lattice-like LO state to sinusoidally-varying
  LO State}
\begin{figure}[htbp]
\centerline{\includegraphics[width=10cm]{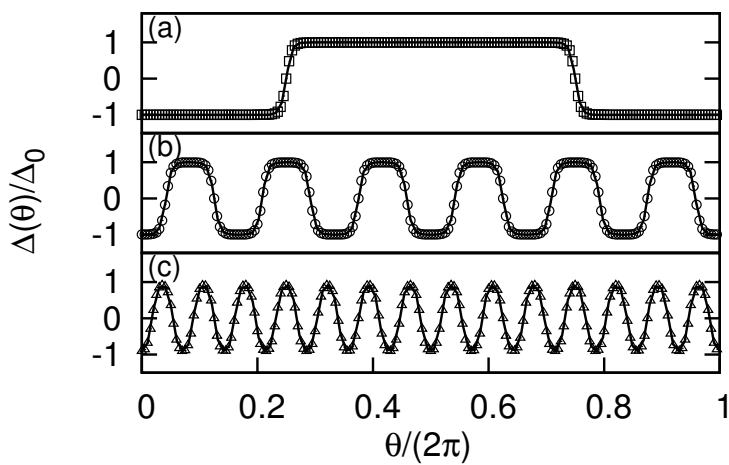}}
\caption[]{\label{fig:1D_order_para} Angle distribution of pairing
  order parameter in an ideal soliton state. The order parameter is
  measured in unit of $\Delta_0$ which is the value of order parameter
  in the uniform state. From top to bottom, total spin imbalance is 1,
  6, and 14.  Open symbols: numerical results; solid lines: fitting
  function $\tilde{\Delta}\tanh(\cos n\theta/\tilde{\xi})$ with two
  parameters $\tilde{\Delta}$ and $\tilde{\xi}$.}
\end{figure}

For equally populated species $N_{\uparrow}=N_{\downarrow}$, the
lowest energy state is obviously the BCS state with uniform pairing
gap. If one spin is flipped from downward to upward, i.e.,
$N_{\uparrow}+1$ up spin and $N_{\downarrow}-1$ down spin, a pair of
soliton and anti-soliton is developed to store these two excess
spins. We define the spin imbalance $n$ to be
$(N_{\uparrow}-N_{\downarrow})/2$ for spin 1/2 particle. A typical
soliton and anti-soliton pair is plotted in
Fig.~(\ref{fig:1D_order_para}a), which is obtained by numerically
solving Eq.~(\ref{eq:1}) in a ring, where we use the angle
$\theta=2\pi x/L$ as the coordinate.  Due to the periodic boundary
condition, a single soliton can not exist freely so that it must
co-exist with an anti-soliton as a pair with the same width $\xi$. We
call these soliton states with each spin per soliton (anti-soliton) as
\emph{ideal soliton} state. Note that since the order parameter is
real, this state is also a kind of LO state. Actually, all the
self-consistent solutions shown in this paper have real order
parameters which minimize the energy, and therefore they are LO state.
In the following sections, we omit ``LO'' for the sake of brevity.

With the increase of the flipped spins, more soliton and anti-soliton
pairs are generated. Thus we get the \emph{soliton lattice} state with
$n$ pairs of soliton and anti-soliton as long as the system is in the
\emph{dilute limit} by which we mean $n\xi\ll 2\pi$, here the soliton
width $\xi$ is measured in unit of the angle. In the dilute limit, the
solitons are well separated from each other, which has two
consequences (i) all the midgap states have zero energy, and (ii) each
soliton or anti-soliton carries exactly one localized spin. According
to these two properties, we distinguish soliton lattice state from the
sinusoidally modulated state, where the spin imbalance $n$ is too
large to satisfy $n\xi < 2\pi$ and solitons overlap considerably with
each other. Then the energy spectrum of the midgap states has a
dispersion described by the Bloch theorem for a periodic lattice. Such
a scenario from soliton lattice to sinusoidally varying state has also
been addressed in Ref.~\cite{buzdin1997} from the viewpoint of GL
theory. The pairing parameter for both states can be described
perfectly by the fitting function $\tilde{\Delta}\tanh(\cos
n\theta/\tilde{\xi})$ \footnote{The soliton lattice pattern of pairing
  parameter can be described by Jacobi elliptic function as done in
  Ref.~\cite{machida1984}, but we do not take that expression for the
  sake of simplicity.}  with $\tilde{\Delta}$ and $\tilde{\xi}$ to be
determined, which is shown in Fig.~\ref{fig:1D_order_para}.

\begin{figure}[htbp]
\centerline{\includegraphics[width=10cm]{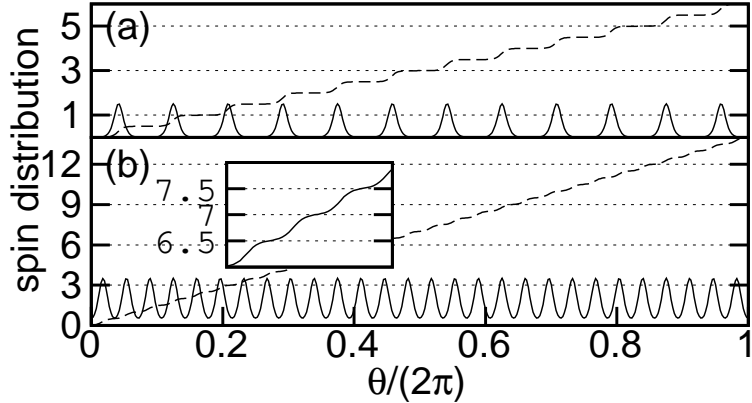}}
\caption[]{\label{fig:1D_spin_dist} Spin distribution in soliton
  lattice state for the system with spin imbalance 6 (a) and 14
  (b). Solid lines: local spin distribution; Dash lines: integrated
  spin distribution. Inset of (b) shows a zoomed figure around a
  plateau.}
\end{figure}

We now introduce two spin distribution functions, local spin
distribution $S_{L}(\theta)=\frac{1}{2}\langle\hat{\psi}^{\dagger}
_{\uparrow}(\theta)\hat{\psi}_{\uparrow}(\theta)-\hat{\psi}^{\dagger}
_{\downarrow}(\theta)\hat{\psi}_{\downarrow}(\theta)\rangle$ as well
as integrated spin distribution $S_I(\theta)$,
\begin{eqnarray}
\label{eq:13}
S_I(\theta)=\int_0^{\theta}S_L(\theta') d\theta'\;.
\end{eqnarray}
As shown in Fig.~\ref{fig:1D_spin_dist}, the localization of spin
density in the soliton lattice state manifests itself in the plateau
features of the function $S_I(\theta)$.  For the sinusoidally
modulated state, the plateaus disappear due to the delocalization of
spins.

\subsection{Deformed Soliton}
\begin{figure}[htbp]
\centerline{\includegraphics[width=10.5cm]{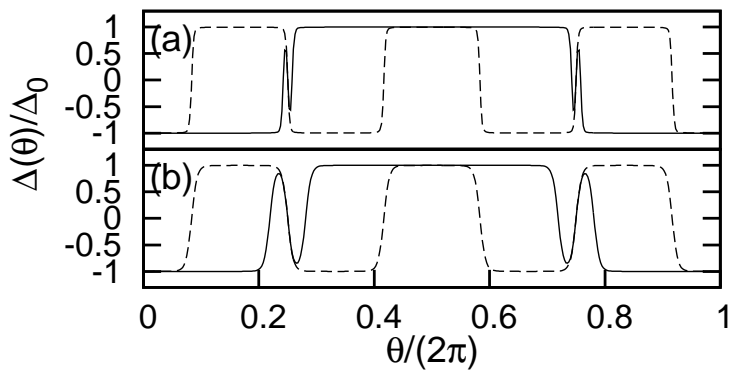}}
\caption[]{\label{fig:1D_ring_order_deformed} Pairing parameter in
  deformed soliton with $\mathcal{Q}=3$ (solid lines) and ideal
  soliton with $\mathcal{Q}=1$(dashed lines). Upper (lower) panel
  corresponds to the strong (weak) pairing interaction $g$.}
\end{figure}

Here we introduce $\mathcal{Q}$ to denote the number of spins per
soliton/antisoliton. In the previous subsections, we focused on the
state with only one spin($\mathcal{Q}=1$) per soliton.  Now we study
the case for $\mathcal{Q}\ge2$, which we call \emph{deformed soliton}
state.  Firstly, let us consider the case for odd $\mathcal{Q}$.  The
order parameter of a deformed soliton state for $\mathcal{Q}=3$ is
plotted in Fig.~\ref{fig:1D_ring_order_deformed}(solid lines), which
corresponds to 6 excess spins in total. Note that these 6 spins can
also be stored in 3 ideal soliton-antisoliton pairs(dashed
lines). Hence, we need to compare their energies numerically. It turns
out that the deformed soliton is energetically favorable for strong
interaction, while the ideal soliton state is preferable for weak
interaction. Note that the deformed soliton found in this article has
$\mathcal{Q}$ nodes in a narrow region. In fact $\mathcal{Q}$ spins
can also be accommodated by a special soliton with only one nodes,
which is described by $\Delta_0\tanh(x/\xi)$ with
$\Delta_0\xi=(\mathcal{Q}+1)/2$(see Ref.\cite{Takayama1979}), however
one can show that this solution is not energetically favored by
comparing its energy and that of the corresponding well separated
multi-soliton state.

In Fig.~\ref{fig:1D_ring_order_deformed}, the upper panel corresponds
to a strong interaction case where the three spins are squeezed in a
very narrow region with width comparable to that of an ideal soliton
$\xi$.  The total width is then estimated to be around $2\xi$, which
is much smaller than the width $6\xi$ for the ideal soliton state.
Thus, one can reasonably believe that the deformed soliton state has
lower energy. If the interaction strength $g$ becomes weaker, as shown
in the lower panel of Fig.~\ref{fig:1D_ring_order_deformed}, the
deformed soliton with $\mathcal{Q}=3$ will inflate and its pattern is
getting close to three ideal solitons. When $g$ becomes weak enough,
the deformed soliton can not be stable, and is transmuted into an
ideal soliton lattice state. The pairing order parameter shown in
Fig.~\ref{fig:1D_ring_order_deformed} can be perfectly fitted with
function $\tilde{\Delta}_0[\tanh(\cos(\theta-\theta_0)/\tilde{\xi}_0)-
\tanh(\cos(\theta)/\tilde{\xi}_0+
\tanh(\cos(\theta+\theta_0)/\tilde{\xi}_0] $ with three parameters
$\tilde{\Delta}_0$, $\tilde{\xi}_0$ and $\tilde{\theta}_0$.

Note that the order parameter has a sign change $(-1)^{\mathcal{Q}}$
after crossing $\mathcal{Q}$ ideal solitons and
antisolitons. Therefore, if $\mathcal{Q}$ is odd, a deformed soliton
can be continuously transmuted into $\mathcal{Q}$ ideal solitons, but
this is not true for even $\mathcal{Q}$ due to the mismatched boundary
condition of $\Delta(x)$. In addition, the energy of a deformed
soliton with even $\mathcal{Q}$ is not energetically favorable in our
numerical calculations. Therefore, we do not need to consider the case
for even $\mathcal{Q}$.

\subsection{Effect of Magnetic Field}
\label{sec:effect-magn-field}
So far we only consider the system with fixed particle number, and
have not included the magnetic field in our analysis. Since the total
spin is a good quantum number, the effect of magnetic field can be
easily estimated by simply adding Zeeman energy
$-\mu_Bh(N_{\uparrow}-N_{\downarrow})$. Obviously, the state with more
excess spins gains magnetic energy, however it is at the cost of the
deformation of pairing gap which loses the condensation
energy. Therefore, the ground state should correspond to an optimized
value of spin imbalance.  

Let $n=(N_{\uparrow}-N_{\downarrow})/2$ be the spin imbalance, and the
corresponding ground state energy be denoted by $E(n)$. The energy of
the BCS state without spin imbalance is thus $E(0)$. Given an external
magnetic field $h$, we then need to find the lowest free energy for
all possible $n$'s, i.e., minimize $E(n)-2n\mu_{B}h$ with respect to
$n$, which leads to an optimal spin imbalance $n_c$.

\begin{figure}[htbp]
\centerline{\includegraphics[width=11cm]{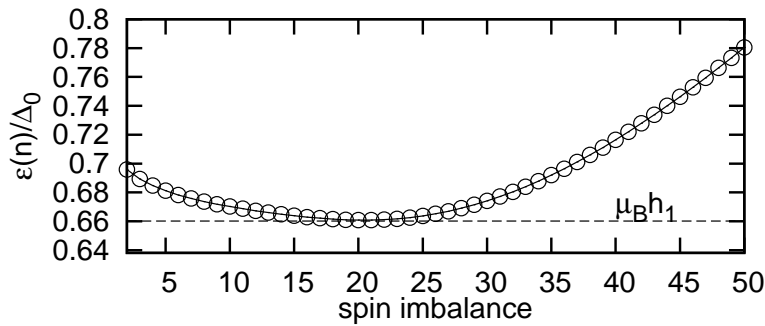}}
\caption[]{\label{fig:1D_energy_per_spin} Average energy per spin
  $\varepsilon(n)$ in Eq.~(\ref{eq:8}) as a function of spin imbalance
  $n$. The dashed line is the first critical magnetic field
  $h_1$.}
\end{figure}

To this purpose, we define the energy cost per spin as
\begin{eqnarray}
\label{eq:8}
\varepsilon(n) \equiv [E(n)-E(0)]/(2n),
\end{eqnarray}
which can also be regarded as the energy cost for creating one
soliton. The numerical data of $\varepsilon(n)$ is plotted in
Fig.~\ref{fig:1D_energy_per_spin}. As $n$ increases, the adjacent
kinks become closer, which enhances the hopping amplitude of spins
between kinks and consequently favors the kinetic energy of spin
transfer. However, at the same time, the pairing gap gets smaller,
which reduces the condensation energy. Thus the interplay between
these two mechanisms leads to the nontrivial pattern of
$\varepsilon(n)$ in Fig.~\ref{fig:1D_energy_per_spin}.

There is a critical value $h_1$ of the magnetic field, below which the
magnetic energy can not support an ideal soliton, and the system
remains in the uniform state.  When $h > h_1$, the sinusoidally
varying state with modulation frequency $n_c$ will become
energetically favorable. $n_c$ can be determined by the minimum of
$2n\varepsilon(n)-2n\mu_Bh$, alternatively, the optimal spin imbalance
$n_c$ should satisfy
\begin{eqnarray}
\label{eq:9}
\left.\frac{\partial(2n(\epsilon(n)-\mu_{B}h))}{\partial n}\right|_{n=n_c} =0
\end{eqnarray}
After a little algebraic analysis of Eq.~(\ref{eq:9}), one can see
that $n_c$ increases as $h$ increases.  The first appeared $n_c$ is
determined by $\epsilon(n_c)=\mu_Bh_1$ which is far from zero as
shown in Fig.~\ref{fig:1D_energy_per_spin} and corresponds to a
sinusoidally modulated  state.

\section{Imbalanced Superfluid State in Annular Disk}

In this section we present our numerical results for imbalanced
superfluid state in narrow annular disk with inner radius $R_1$ and
outer radius $R_2$. The radial width $R_2-R_1$ is small enough to
avoid the modulation of order parameter along the radial direction.
In the numerical calculation, we use the ratio
$\rho\equiv(R_2-R_1)/R_1$ to characterize the geometry of annular
disk.  Since $g$ has the dimension of
[energy]$\cdot$[length]$^2$, a dimensionless quantity
$\tilde{g}\equiv g/(\pi(R_2^2-R_1^2)\mu)$ is introduced to represent
the interaction strength. The BdG equation is solved in momentum
space. Most of the results in this section are based upon the
diagonalization of Hamiltonian in a Hilbert space with dimensionality
3500 and 11 radial modes involved.

\subsection{Fixing Particle Number $N_{\uparrow}$ and
  $N_{\downarrow}$}

\subsubsection{ Ideal Domain Wall }
For small spin imbalance, one should get domain walls as an analog of
solitons in 1D case, and the excess spins are attached to the domain
walls. It is natural to ask what is the optimal number($\mathcal{Q}$)
of spins per domain wall. To answer this question, we first consider
an ideal geometry, i.e., a narrow strip with periodic boundary
condition in both $x$ and $y$ directions, but with length $L_x\gg
L_y$.

This simplified model reads
\begin{eqnarray}
\label{eq:10}
\hat{H}&=&\int dxdy [\hat{\psi}^{\dagger}_{\alpha}\left(
  \frac{\hat{\vec{p}}^2}{2m}-\mu_{\alpha} \right) \hat{\psi}_{\alpha} \nonumber\\
& +& \Delta(x,y)\hat{\psi}^{\dagger}_{\uparrow}\hat{\psi}^{\dagger}_{\downarrow} +
\Delta^{*}(x,y) \hat{\psi}_{\downarrow}\hat{\psi}_{\uparrow}
-\frac{|\Delta(x,y)|^2}{g}] \nonumber\\
\Delta(x,y)&=&g \left\langle \hat{\psi}_{\downarrow}
  \hat{\psi}_{\uparrow} \right\rangle\;.
\end{eqnarray}
The ideal domain wall pattern of $\Delta(x,y)$ is independent of $y$,
and has the form $\Delta(x,y) =\Delta_0 \tanh(x/\xi_0)$
which implies the pairing momenta in $y$ direction are always $q$ and
$-q$.  The Hamiltonian in Eq.~(\ref{eq:10}) can be divided into many
1D branches with respect to the discrete momenta
$q=(2\pi/L_y)\times\mbox{integer}$ in $y$ direction,
\begin{eqnarray}
\label{eq:12}
\hat{H}_q  &\sim&\int dx [\hat{\psi}^{\dagger}_{q,\uparrow}(x)\left(
  \frac{\hat{p}_x^2}{2m}-\mu_{q\uparrow} \right) \hat{\psi}_{q,\uparrow}(x)
\nonumber\\
&&\hspace{0.7cm}
+\hat{\psi}^{\dagger}_{-q,\downarrow}(x)\left(
 \frac{\hat{p}_x^2}{2m}-\mu_{-q\downarrow} \right) \hat{\psi}_{-q,\downarrow}(x)
\nonumber\\
&+& \Delta\hat{\psi}^{\dagger}_{q,\uparrow}(x)\hat{\psi}^{\dagger}_{-q,\downarrow}(x) +
\Delta^{*}\hat{\psi}_{-q,\downarrow}(x)\hat{\psi}_{q,\uparrow} (x)]\;.
\end{eqnarray}
Note that $\Delta(x)$ is contributed from all 1D branches, and the
$q$-dependent chemical potential reads
$\mu_{q\alpha}=\mu_{\alpha}-(\hbar q)^2/(2m)$, which are determined by
the particle numbers $N_\alpha$. Each $q$-mode with $\mu_{q}>0$ can
accommodate one spin per soliton. Therefore, we can estimate the
optimal spin filling $\mathcal{Q}$ of each ideal domain wall to be the
number of $q$-modes buried under the FS. The optimal filling for the
annular disk with open boundary condition in the radial direction can
also be estimated similarly by counting the number of energy modes
under the FS.

\begin{figure}[htbp]
\centerline{\includegraphics[width=10cm]{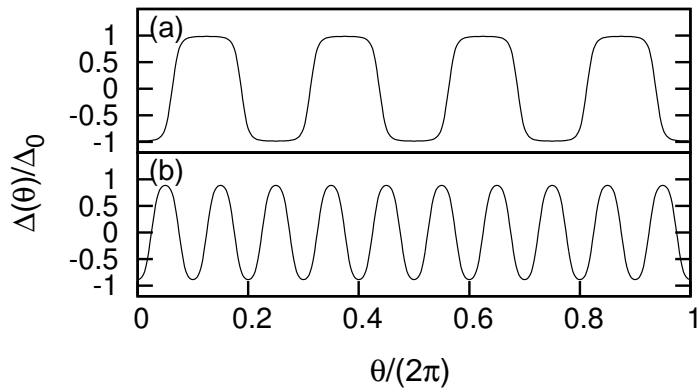}}
\caption[]{\label{fig:2D_order_para} Angle dependence of pairing order
  parameter at radius $(R_1+R_2)/2$. (a): Domain wall lattice state
  with total spin $28$; (b): sinusoidally-varying LO state with
  total spin $70$. The optimal filling per domain wall is
  $\mathcal{Q}=7$.  The system parameter $\rho=0.4$.}
\end{figure}

\begin{figure}[htbp]
\centerline{\includegraphics[width=9cm]{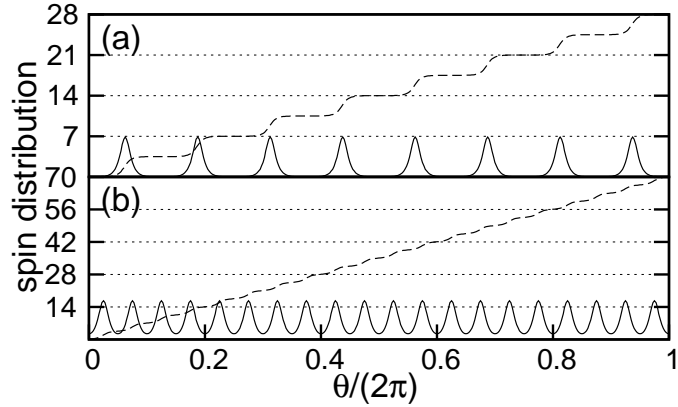}}
\caption[]{\label{fig:2D_spin_dist} Spin distribution in domain wall
  lattice state with spin imbalance 28 (upper panel) and in
  sinusoidally-varying state with spin imbalance 70 (lower panel).
  Dashed lines: integrated spin distribution $S_I(\theta)$, and solid
  lines: local spin distribution $S_L(\theta)$. The system parameters
  are the same as in Fig. \ref{fig:2D_order_para}.}
\end{figure}

\begin{figure}[htbp]
\centerline{\includegraphics[width=10cm]{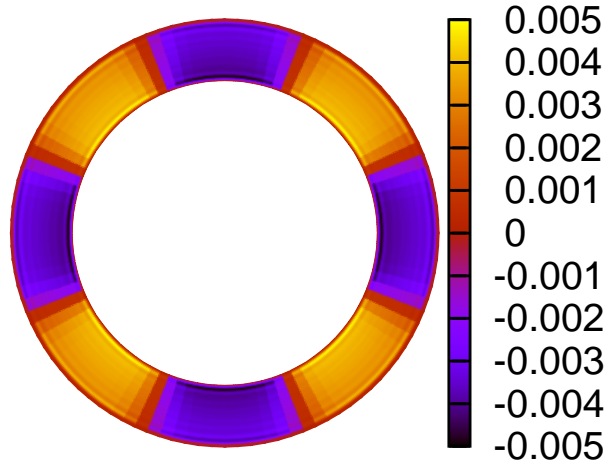}}
\caption[]{\label{fig:3D_order} (Color online.)Contour plot of order
  parameter. The excess spin equals to $28$ and the optimal filling in
  this case is $\mathcal{Q}=7$, hence four pairs of domain walls are
  needed to store these excess spins. the system parameters are the
  same as in Fig. \ref{fig:2D_order_para}.}
\end{figure}

Similar to the 1D ring, one expects a crossover from an ideal domain
wall like LO state to the sinusoidally-varying LO state with
increasing spin imbalance in the weak interaction case.  Since
$\Delta(r,\theta)$ now depends on $r$, we plot the angle dependence of
$\Delta(r,\theta)$ at radius $r=(R_1+R_2)/2$ in
Fig.~\ref{fig:2D_order_para}. The full spatial dependence of
$\Delta(r,\theta)$ is plotted in 2D contour in
Fig.~\ref{fig:3D_order}, where one can find its radial dependence is
nearly uniform.  The spin density $s(r,\theta)$ is also a function of
$r$ and $\theta$. By integrating $s(r,\theta)$ over $r$, we can define
angle dependent local spin distribution $S_L(\theta)$, and angle
dependent integrated spin distribution $S_{I}(\theta)$, as following,
\begin{eqnarray}
\label{eq:14}
&&S_{L}(\theta) = \int_{R_1}^{R_2}rdr s(r,\theta) \nonumber\\
&&S_{I}(\theta) = \int_0^{\theta} S_L(\theta) d\theta'\;.
\end{eqnarray}
$S_L$ and $S_I$ are plotted as functions of $\theta$ in
Fig.~\ref{fig:2D_spin_dist}, which shows clearly that the spin
distribution are localized in the domain wall state, and delocalized
in the sinusoidally-varying LO state.

\subsubsection{Deformed Domain Wall and Phase Separation}
\begin{figure}[htbp]
\centerline{\includegraphics[width=10cm]{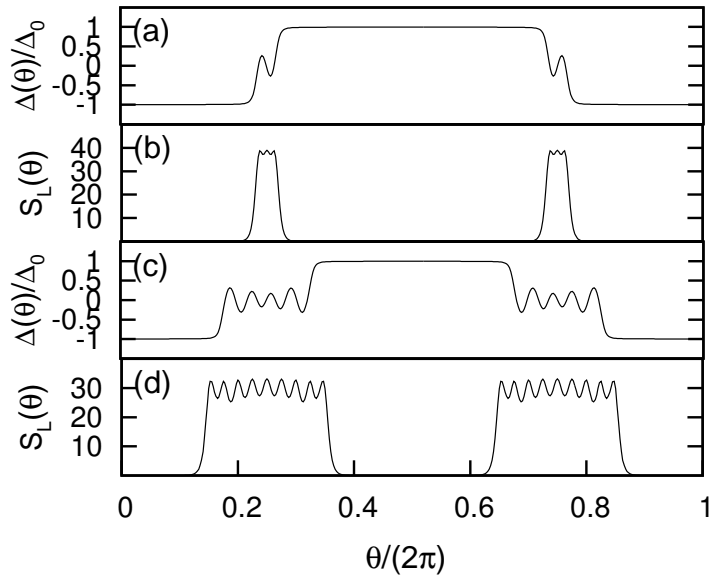}}
\caption[]{\label{fig:2D_deformed} Deformed domain wall [(a) and (b)]
  and phase separation [(c) and (d)] solutions.  We plot the order
  parameter in (a) and (c), and spin distribution in (b) and (d).  The
  spin imbalance is 21 for (a) and (b), and 77 for (c) and (d). The
  optimal spin filling $\mathcal{Q}=7$. The interaction strength is
  $\tilde{g} \sim 6.9\times 10^{-4}$.}
\end{figure}

As in the 1D ring, we also encounter the deformed domain wall state,
for which there can be more spins than the optimal filling
$\mathcal{Q}$ squeezed in one domain wall. These deformed domain wall
states are stabilized by the \emph{strong} pairing interaction. We
plot the order parameter $\Delta(\theta,r)$ and local spin
distribution $S_L(\theta)$ in Fig.~\ref{fig:2D_deformed}, which shows
that when the spin number exceeds the optimal filling, instead of
creating more ideal domain walls, the spin polarized regions are
simply enlarged.  Note that in the polarized region there is still a
small pairing oscillation like a mini sinusoidally-varying LO state in
order to further lower the potential energy. These deformed domain
wall states (see Fig.~\ref{fig:2D_deformed}c) are then considered as a
kind of \emph{phase separation} state, where the polarized normal
state with small fluctuating order parameter is separated with the
fully pairing phase without spin imbalance.

\subsubsection{Quasiparticle Density of States}
We compute the quasiparticle density of states (DOS) in this section
which can describe the low energy excitations of various ground
states. In our calculation the Zeeman energy is not included, which
corresponds to the situation with fixed particle numbers.  We find
that, for the domain wall lattice state there is a zero energy peak in
the quasiparticle DOS. As the spin imbalance is increasing, the number
of domain walls grows and it results in the enhancement of the zero
energy peak. These zero modes can also be understood from the aspect
of Andreev reflection\cite{vorontsov2005}, since the $\pi$-phase
difference between two superfluids allows an Andreev bound state
located at the domain walls. In the phase separation case, the system
mimics a superconductor-normal metal-superconductor junction. By
increasing the width of normal metal region, more Andreev resonance
states enter into the gap with nonzero energy. These energy levels
then distribute evenly in the gap, which form a flat quasiparticle DOS
in the superconducting gap.

\begin{figure}[htbp]
\centerline{\includegraphics[width=10cm]{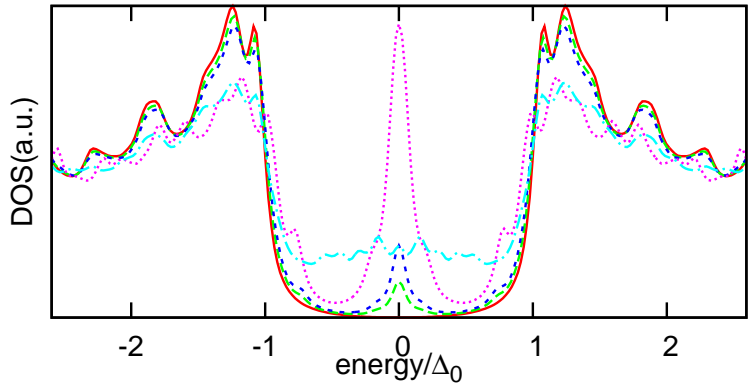}}
\caption[]{\label{fig:dos} (Color online.) Quasiparticle density of
  state for different ground states. The Zeeman energy is not included
  in this figure. The red solid line is for the uniform BCS state, the
  green long dashed line and the blue short dashed line are for the
  domain wall lattice states, the dotted pink line is for the
  sinusoidally modulated LO state, and the cyan dot-dashed line
  corresponds to the phase separation state.}
\end{figure}

The above theoretical analysis is in good agreement with the numerical
results presented in Fig.~\ref{fig:dos}. The DOS of BCS state is zero
in the gap. When increasing the spin imbalance in the ideal domain
wall lattice state, the peak of DOS centered around zero becomes
higher, which means more domain walls are created. Whereas in the case
of phase separation, the DOS in the gap is quite flat due to the
presence of polarized normal state.

\subsection{Fixing Chemical Potentials $\mu_{\uparrow}$ and $\mu_{\downarrow}$}
\begin{figure}[htbp]
\centerline{\includegraphics[width=10cm]{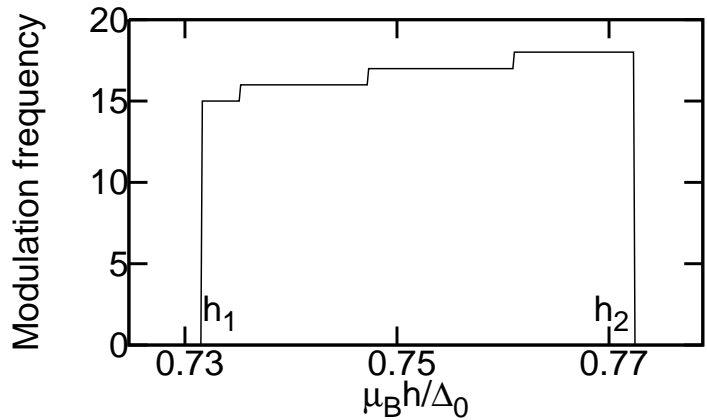}}
\caption[]{\label{fig:period-h} Frequency of pairing modulation as a
  function of magnetic field. We set $\rho=0.2$, and
  $\tilde{g}=5\times 10^{-4}$. $\mu_B$ is the Bohr magneton, and
  g-factor of electron is taken as 2.}
\end{figure}

In this subsection, we show the numerical results in the grand
canonical ensemble with fixed chemical potentials. For weak magnetic
field($2\mu_{B}h=\mu_{\uparrow}-\mu_{\downarrow}$), the Zeeman energy
is not enough to break the $s$-wave Cooper pairs, so the system
retains the uniform BCS state. Until the magnetic field $h$ exceeds
its first critical value $h_1$, the sinusoidally-varying LO state
emerges. As the magnetic field is further increased, the modulation
frequency of the order parameter becomes larger while its magnitude is
reduced, until the system enters into the normal state at the second
critical magnetic field $h_2$. We plot modulation frequency as a
function of $h$ in Fig.~\ref{fig:period-h}, where one can find
plateaus, since there should be integral pairs of domain walls in a
ring geometry.

The phase separation(deformed domain wall) state can not be a ground
state in the homogeneous magnetic field, except at the critical value
$h_1$ of magnetic field. Furthermore, unlike the case of fixing
particle number, there is no continuous crossover from domain wall
state to the sinusoidally-varying state. The onset frequency at the
critical magnetic field $h_1$ is finite and large enough to form a
sinusoidally-varying LO state. The reason is that, to sustain a single
domain wall, its magnetic energy gain must fully compensate the energy
loss due to the deformation of pairing gap. In such a case there can
be more domain walls.  However the overlap of domain walls suppresses
the pairing gap inevitably, which causes the loss of the condensate
energy(see sec.~\ref{sec:effect-magn-field}). At the balance point of
these two processes, sinusoidally-varying state shows up accompanied
with delocalized spins.

\section{Conclusion}

We have investigated the imbalanced superfluid state in annular disks
and 1D rings by solving the BdG equation in the momentum space at zero
temperature. A key issue of imbalance superfluid is how to accommodate
the excess spins by adjusting the pairing gap $\Delta(\vec{r})$. There
are several possibilities, e.g. the LO state with periodically
oscillated order parameter and the phase separation state. We show
that these states are stable under different conditions.

Firstly, we have studied the case with fixed fermion numbers, which
may be relevant to cold atom systems. For low spin imbalance (still
larger than the optimal spin filling $\mathcal{Q}$ per domain wall),
the solitons in 1D and domain walls in 2D are the ground states. The
number of spins localized at each soliton or domain wall is
quantized. When increasing spin imbalance, more and more domain
walls(solitons) occur and overlap with each other, and the
sinusoidally-varying state emerges with delocalized spins. These two
states are distinguished in this paper due to their different spin
distribution. There should be a crossover between them if one tunes
the spin imbalance continuously. The above argument is valid for weak
interactions, whereas for strong interactions, the phase separation is
the possible ground state, in which only the area of normal polarized
state varies with the spin imbalance. This may serve as a criteria to
distinguish the phase separation state and the periodically
oscillating LO state.

Secondly, we have addressed the case of fixing chemical potential
$\mu$ and magnetic field $h$, which may be relevant to heavy fermion
superconductors interacting with an external magnetic field via the
Zeeman term. There are two critical magnetic fields $h_1$ and $h_2$,
which correspond to the transition from uniform BCS state to the
sinusoidally-varying state, and from the sinusoidally-varying state to
the normal state, respectively. It is stressed that the modulation
frequency of pairing gap at $h_{1}$ is quite large and the spin is
delocalized, which characterizes a typical sinusoidally-varying state.

\ack

F.Y. would like to thank T. Li, H. Zhai and Z. B. Su for many
stimulating discussions. This work was supported by RGC grants in
HKSAR, the National Natural Science Foundation of China (Grants
No. 1054700, No. 10874032) and the State Key Programs of China (Grant
No. 2009CB929204). Y.C. acknowledges the program for Professor of
Special Appointment (Eastern Scholar) at Shanghai Institutions of
Higher Learning.

\section*{Reference}
\bibliographystyle{unsrt}
\bibliography{fflo}
\end{document}